\documentstyle[vsolj01,graphicx,natbib]{article}

\RequirePackage[T1]{fontenc}

\def\cite{\citealt}

\def\Shibataprep{M. Shibata et al. in preparation}

\begin{document}

\title{On the nature of embedded precursors in long outbursts of}
\title{SS Cyg stars as inferred from observations of the IW And star ST Cha}

\author{Taichi Kato$^1$, Franz-Josef Hambsch$^{2,3,4}$}
\author{$^1$ Department of Astronomy, Kyoto University,
       Sakyo-ku, Kyoto 606-8502, Japan}
\email{tkato@kusastro.kyoto-u.ac.jp}
\author{$^2$ Groupe Europ\'een d'Observations Stellaires (GEOS),
     23 Parc de Levesville, 28300 Bailleau l'Ev\^eque, France}
\author{$^3$ Bundesdeutsche Arbeitsgemeinschaft f\"ur Ver\"anderliche
     Sterne (BAV), Munsterdamm 90, 12169 Berlin, Germany}
\author{$^4$ Vereniging Voor Sterrenkunde (VVS), Oostmeers 122 C,
     8000 Brugge, Belgium}

\begin{abstract}
We observed the IW And-type dwarf nova ST Cha and found
that standstills were terminated by brightening at a constant
brightness level during standstills.  This finding is
not consistent with a model of IW~And-type dwarf novae
assuming repeated enhancements of the mass-transfer rate
from the secondary.  We found that one outburst in ST Cha
had a shoulder during the rising branch at the same
level in which standstills were terminated by brightening.
This phenomenon is very similar to what are called
``embedded precursors'' in SS~Cyg stars.  We propose
that these embedded precursors in both SS Cyg stars
and the IW And star ST~Cha occur when the disk reaches
the tidal truncation radius.  If this is the case,
precursors in SS Cyg stars and SU~UMa stars are
different in origin on the contrary to the idea
suggested by \citet{can12ugemLC}.
\end{abstract}

\section{Introduction}

   Superoutbursts of SU UMa-type dwarf novae (SU UMa stars)
are considered to arise from when the disk reaches
the 3:1 resonance, which triggers tidal instability
(\cite{whi88tidal}; \cite{hir90SHexcess};
\cite{lub91SHa}).  Although this picture of
the thermal tidal instability (TTI) model (\cite{osa89suuma};
\cite{osa96review}) was once challenged by
\citet{sma13negSH}, analysis of the Kepler high-precision
observation of V1504 Cyg and V344 Lyr concluded
that the TTI model is the only viable model for ordinary
SU UMa-type dwarf novae (\cite{osa13v1504cygKepler};
\cite{osa13v344lyrv1504cyg};
\cite{osa14v1504cygv344lyrpaper3}).

   On the other hand, it has been shown that
light curves of SS Cyg stars, in which
the 3:1 resonance is unlikely to occur, show
a complex feature during their long outbursts
using high-precision photometry \citep{can12ugemLC}.
These outbursts start with a precursor-like part
as in superoubursts of SU UMa-type dwarf novae,
and \citet{can12ugemLC} called the phenomenon
``embedded precursor''.  Following this discovery,
\citet{can12ugemLC} supported van Paradijs' traditional
idea \citep{vanpar83superoutburst} that long outbursts 
in dwarf novae above the period gap and superoutbursts
in systems below the period gap constitute a unified class.
In this picture, long outbursts in SS~Cyg stars and
superoutbursts in SU UMa stars are essentially the same
and superhumps in SU UMa stars appear as the result
of the long durations of superoutbursts enabling
the 3:1 resonance to grow.  Whether this traditional
picture or Osaki's TTI model is correct has long been
discussed (see \cite{osa13v1504cygKepler}).
To solve this issue, it is necessary to understand
whether ``embedded precursors'' in SS Cyg stars
and precursors in SU~UMa-type superoutbursts
are the same phenomenon or not.

   In the meantime, a new subclass of dwarf novae
IW And stars (\cite{sim11zcamcamp1}; \cite{kat19iwandtype})
have been recognized and have been receiving attention
both from observational and theoretical sides
(\cite{sim11zcamcamp1}; \cite{szk13iwandv513cas};
\cite{ham14zcam}; \cite{kat20imeri};
\cite{kim20kic9406652}; \cite{kim20iwandmodel};
\cite{kat20bccas}; \cite{kim21kic9406652};
\cite{lee21hopup}; \cite{kat21bocet}).

   ST Cha is one of IW And stars (\cite{sim14stchabpcra};
\cite{kat19iwandtype}).  One of the authors (FJH) obtained
time-series photometry between 2015 and 2021.
During the 2020--2021 season, this object showed
typical IW~And-type behavior.

\section{Results and Discussion}

\begin{figure*}
  \begin{center}
    \includegraphics[width=16cm]{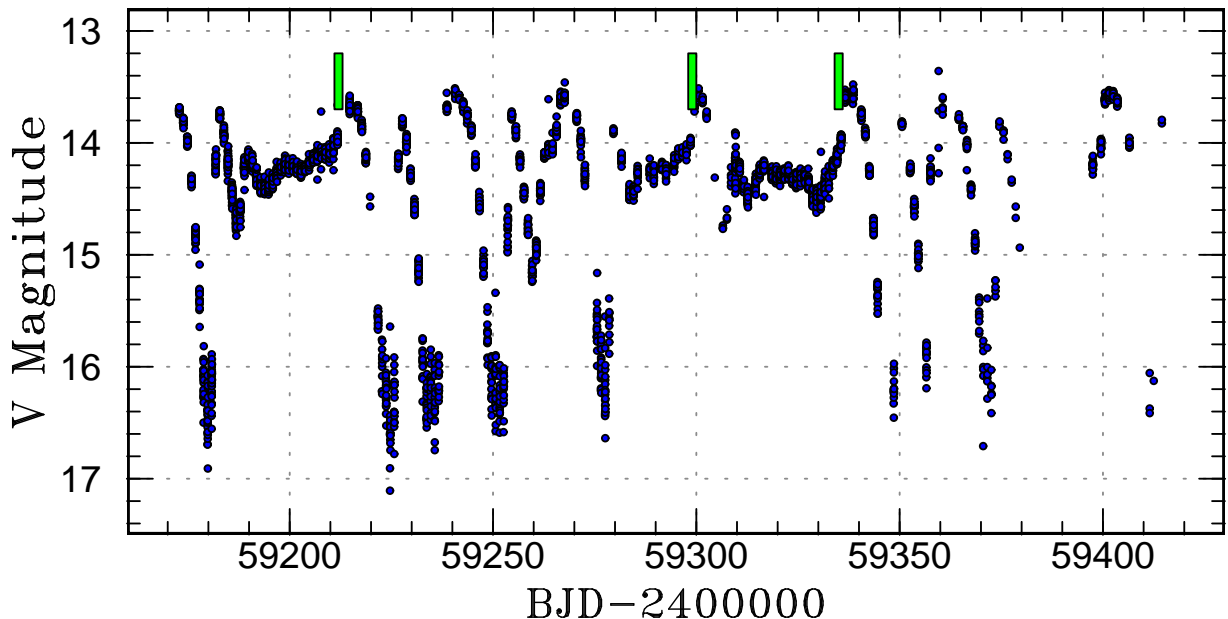}
  \end{center}
  \caption{$V$-band light curve of ST Cha during
  the 2020--2021 season.  IW And-type cycles were repeated
  three times.  The vertical green bars represent
  the epochs when standstills were terminated by
  brightening.}
  \label{fig:stcha2020}
\end{figure*}

   The $V$-band light curve in the 2020--2021 season
is shown in figure \ref{fig:stcha2020}.
In three intervals BJD 2459190--2459212, 2459283--2459299
and 2459309--2459336, this object showed standstills
terminated by brightening and subsequent fading (dip).
The second standstill was followed by brigthening,
a small dip, and immediately another standstill.
This sequence standstill--brigthening--dip
is the basic definition of IW And stars
\citep{kat19iwandtype} [The depths of dips are usually
variable, and in some cases they are completely missing.
This type of variation is referred to as ``heartbeat''.
See the case of HO Pup in \citet{kim20iwandmodel}].
During the first two standstills, the object slowly
brightened suggesting that the disk mass was building up
during these standstills.  The third standstill was
more complex with temporary fading (BJD 2459328--2459332),
but followed by brightening.
We performed a period analysis using these three
standstills without yielding a statistically significant
period.  This may indicate the lack of an orbital or
superhump signal in this system or it may be simply
due to the lack of sensitivity of a period longer than
0.2~d, which was limited to the durations of
nightly runs.

\begin{figure*}
  \begin{center}
    \includegraphics[width=16cm]{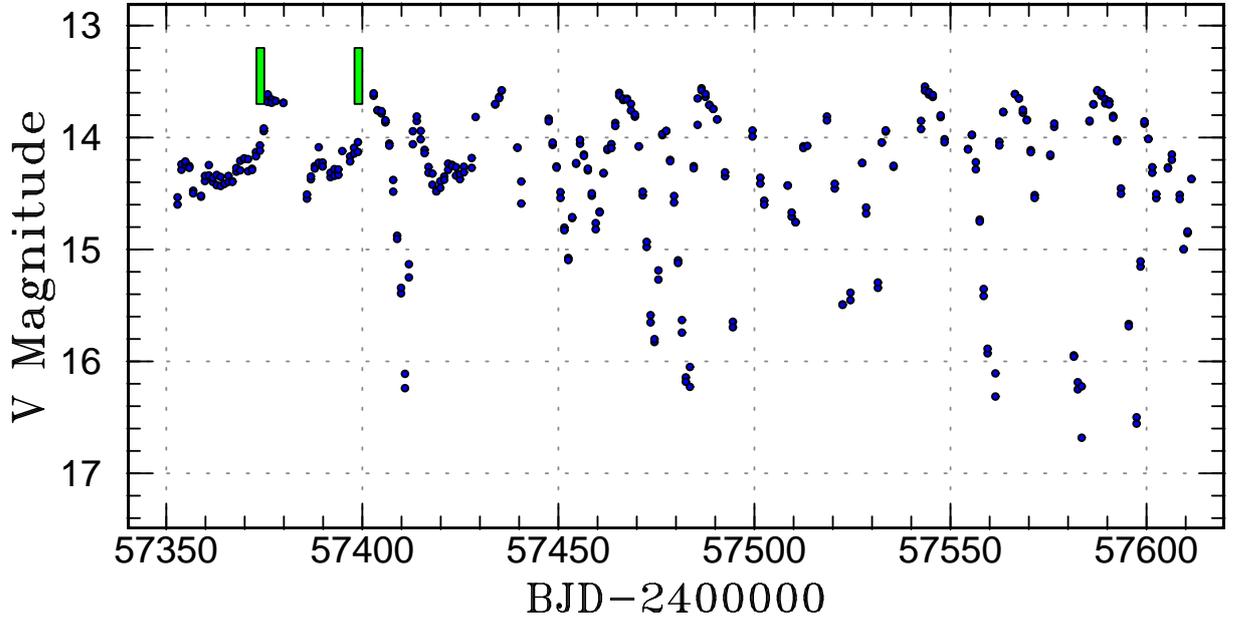}
  \end{center}
  \caption{$V$-band light curve of ST Cha during
  the 2015--2016 season.  IW And-type cycles were repeated
  at least twice early in the season.
  The vertical green bars represent
  the epochs when standstills were terminated by
  brightening.}
  \label{fig:stcha2015}
\end{figure*}

   We found the constancy ($V$=14.0) of brightness
when these standstills started rising to brightening
which terminated these standstills.
Although the IW And-type phenomenon lasted shorter
than in the 2020--2021 season, the same value was
obtained from observations of the 2015--2016 season
(figure \ref{fig:stcha2015}).
The same phenomenon was also observed in 
BO Cet \citep{kat21bocet}.
\citet{ham14zcam} suggested a model of IW And stars
in which repeated enhancements of the mass-transfer
rate from the secondary cause brightening at the end of
standstills.  In this model (although Hameury and Lasota
themselves did not consider it very realistic),
timings of these enhancements of
the mass-transfer rate is determined by the secondary,
not by the state of the disk, and the disk luminosity
at the onset of brightening is not expected to be
constant.  Our observations, however, suggest that
brightening occurs when the luminosity of the disk
reaches a certain level (such as determined by
the disk mass or radius).

\begin{figure*}
  \begin{center}
    \includegraphics[width=16cm]{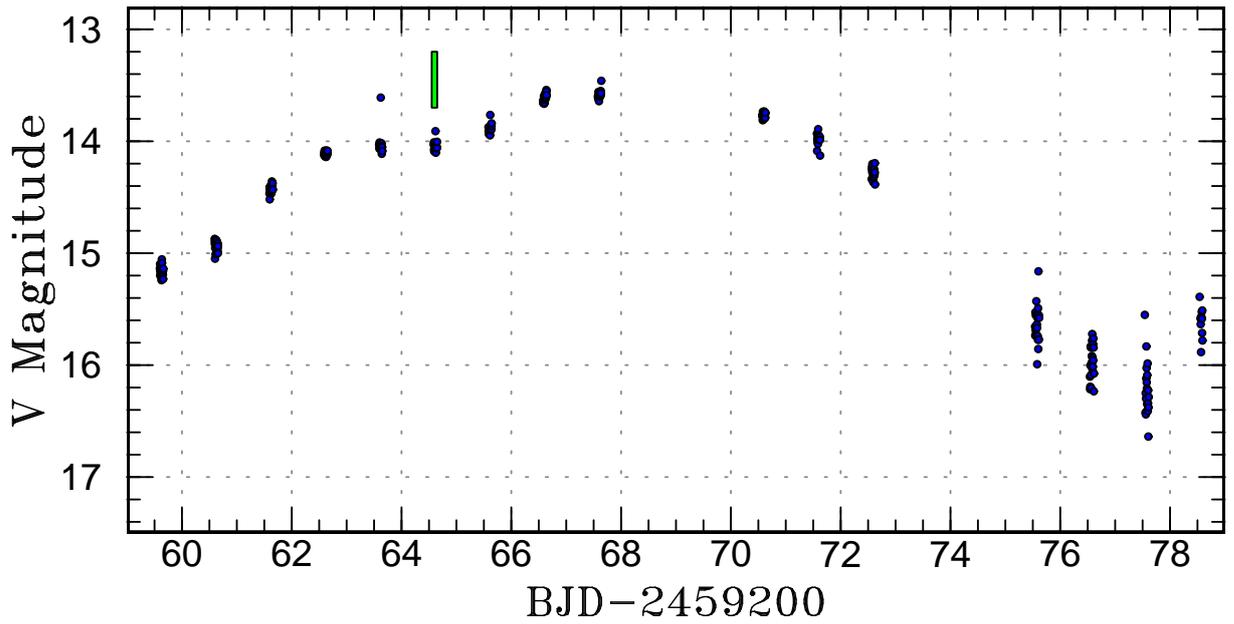}
  \end{center}
  \caption{Outburst of ST Cha with a shoulder
  (embedded precursor).
  The vertical green bar represents the end of
  the shoulder.}
  \label{fig:stchashoulder}
\end{figure*}

   There has been direct observational determination of
the disk radius by \Shibataprep \, which indicates that
the disk radius in IW And-type standstills grows with
time and that standstills are terminated by brightening
when the disk radius eventually reaches
the tidal truncation radius.  The constancy of brightness
at which standstills are terminated by brightening
in BO Cet and ST Cha are also supportive of this idea.
The constant brightness when standstills are terminated
by brightening can be considered as the state
when the disk reaches the tidal truncation radius.
In ST Cha, we found an outburst with an apparent
shoulder or ``embedded precursor''
(figure \ref{fig:stchashoulder}) similar to the ones
observed in SS Cyg stars by \citet{can12ugemLC}.
During this outburst, the shoulder had the same
brighteness as stated above.

   Considering the similarity of embedded precursors
in SS Cyg stars with the phenomenon in ST Cha,
we propose that embedded precursors in SS Cyg stars
refer to a phenomenon in which the disk in these
systems reaches the tidal truncation radius.
If this is correct, the causes of precursors
in SS Cyg stars and SU UMa stars are different since
the tidal truncation radius is outside the radius
of the 3:1 resonance in SU UMa stars.

\section*{Acknowledgments}

This work was supported by JSPS KAKENHI Grant Number 21K03616.

\end{document}